\documentclass[conference]{IEEEtran}

\usepackage{amsmath,amssymb,amsfonts}
\usepackage{graphicx}
\usepackage{booktabs}
\usepackage{algorithm}
\usepackage{algorithmic}
\usepackage{hyperref}
\usepackage{cite}
\usepackage{float}
\usepackage{graphicx}
\usepackage{placeins}

\usepackage{booktabs}
\usepackage[most]{tcolorbox}
\usepackage{xcolor}

\tcbset{
  promptbox/.style={
    colback=gray!8,
    colframe=gray!60,
    boxrule=0.5pt,
    arc=1pt,
    left=6pt,right=6pt,top=4pt,bottom=4pt,
    enhanced,
    fontupper=\ttfamily\footnotesize,
    fonttitle=\bfseries\footnotesize,
  }
}

\begin{document}

\title{NeuroMambaLLM: Dynamic Graph Learning of fMRI Functional Connectivity in Autistic Brains Using Mamba and Language Model Reasoning}

\author{
Yasaman Torabi\textsuperscript{1,*,\textdagger},
Parsa Razmara\textsuperscript{2,*},
Hamed Ajorlou\textsuperscript{3},
Bardia Baraeinejad\textsuperscript{4}
}

\maketitle

\begingroup
\renewcommand\thefootnote{}
\footnotetext{
\begin{list}{}{\setlength{\leftmargin}{0pt}\setlength{\itemindent}{0pt}\setlength{\labelsep}{0pt}\setlength{\labelwidth}{0pt}}

\item\textsuperscript{\textdagger} Corresponding author: torabiy@mcmaster.ca\\
\textsuperscript{*} Equal contribution authors\\
\textsuperscript{1} Y. T. is with the Department of Electrical and Computer Engineering, McMaster University, Hamilton, ON L8S~4L8, Canada.\\
\textsuperscript{2} P. R. is with the Department of Biomedical Engineering, University of Southern California, Los Angeles, CA 90089, USA.\\
\textsuperscript{3} H. A. is with the Department of Electrical and Computer Engineering, University of Rochester, Rochester, NY 14627, USA.\\
\textsuperscript{4} B. B. is the CEO at the BIOSEN Group, No. 15, Nafisi Street, Ekbatan Town, Tehran 1393774535, Iran.\\
\end{list}%

}
\endgroup

\begin{abstract}
Large Language Models (LLMs) have demonstrated strong semantic reasoning across multimodal domains. However, their integration with graph-based models of brain connectivity remains limited. In addition, most existing fMRI analysis methods rely on static Functional Connectivity (FC) representations, which obscure transient neural dynamics critical for neurodevelopmental disorders such as autism. Recent state-space approaches, including Mamba, model temporal structure efficiently but are typically used as standalone feature extractors without explicit high-level reasoning. We propose \textbf{NeuroMambaLLM}, an end-to-end framework that integrates dynamic latent graph learning and selective state-space temporal modelling with LLMs. The proposed method learns the functional connectivity dynamically from raw Blood-Oxygen-Level-Dependent (BOLD) time series, replacing fixed correlation graphs with adaptive latent connectivity while suppressing motion-related artifacts and capturing long-range temporal dependencies. The resulting dynamic brain representations are projected into the embedding space of an LLM, where the base language model remains frozen and lightweight low-rank adaptation (LoRA) modules are trained for parameter-efficient alignment. This design enables the LLM to perform both diagnostic classification and language-based reasoning, allowing it to analyze dynamic fMRI patterns and generate clinically meaningful textual reports. Experiments on the Autism Brain Imaging Data Exchange (ABIDE) dataset demonstrate up to 72.12\% test accuracy, comparable to recent state-of-the-art approaches.
\end{abstract}

\begin{IEEEkeywords}
Neuroimaging, Large Language Models, Graph Neural Networks, Mamba, fMRI, Functional Connectivity
\end{IEEEkeywords}

\section{Introduction}
\label{sec:intro}
Functional Magnetic Resonance Imaging (fMRI) is a cornerstone of modern neuroimaging, providing a non-invasive view of the brain’s functional organization. However, fMRI data is inherently high-dimensional, noisy, and governed by non-linear spatiotemporal dynamics that challenge traditional analysis \cite{glover2011fmri}. Prior neuroimaging studies highlight the sensitivity of fMRI signals to motion, preprocessing confounds, and acquisition variability, while spatial regularization and systematic clinical analyses underscore the importance of reliable neuroimaging findings and translational biomarkers for clinical decision-making \cite{power2012spurious, razmara2024advancements, razmara2025feasibility,momeni2025incidence}. This complexity is especially pronounced in neurodevelopmental conditions such as Autism Spectrum Disorder (ASD). Recent studies show that ASD involves not only structural connectivity deficits, but also abnormal dynamics, including transient hyper-connectivity, unstable state transitions, and short-timescale temporal variability \cite{vidaurre2021brain, supekar2022aberrant, wang2023dynamic}. As a result, the brain is better modelled as a time-varying graph with continuously evolving functional associations \cite{cui2022braingnn, asadi2025graph}. Beyond classification accuracy, clinical use of fMRI-based models requires interpretability and contextual reasoning over neural dynamics. Large Language Models (LLMs) enable structured reasoning by mapping learned brain features into a semantic space that captures relationships, temporal dependencies, and diagnostic context \cite{torabi2023nmf_bss, torabi2025llm_nmf, torabi2025chem_nmf, Torabi2026LingoNMF }. Recent benchmark studies evaluating structured visual reasoning demonstrate that pretrained models encode non-trivial compositional capabilities, and that downstream reasoning performance reveals behaviors not captured by raw accuracy metrics \cite{hudson2019gqa}. Recent multimodal zero-shot and training-free time-series reasoning frameworks demonstrate that pretrained LLMs can generalize across domains when aligned with structured temporal representations \cite{radford2021learning, aryashad2025defogging_vlms}. Integrating LLM reasoning with dynamic brain graphs moves models beyond black-box predictions toward clinically meaningful explanations grounded in temporally specific patterns. Despite this, current Brain–LLM approaches rely heavily on static Functional Connectivity (sFC). Recent frameworks such as FCN-LLM \cite{fcnllm2025} and BrainLLM \cite{zhou2024brainllm} typically summarize the full scan into a single Pearson correlation matrix, collapsing temporal variability and treating the dynamic brain as static \cite{lurie2020questions}. Temporal deep learning models, including Spatiotemporal Graph Convolutional Networks (ST-GCNs) \cite{gadgil2020spatio, kim2021learning}, spatiotemporal encoder-decoder architectures \cite{tran2015learning, yang2020robust},  and Transformers \cite{kan2022brain}, partially address this issue but suffer from scalability limits. In particular, the quadratic cost of self-attention makes end-to-end modelling of long fMRI sequences impractical without aggressive downsampling or windowing, which can introduce bias and artifacts \cite{zhu2024vision}.

\begin{figure}[]
    \centering
    \includegraphics[width=0.48\textwidth,
        trim=0 0 0 0,
        clip
    ]{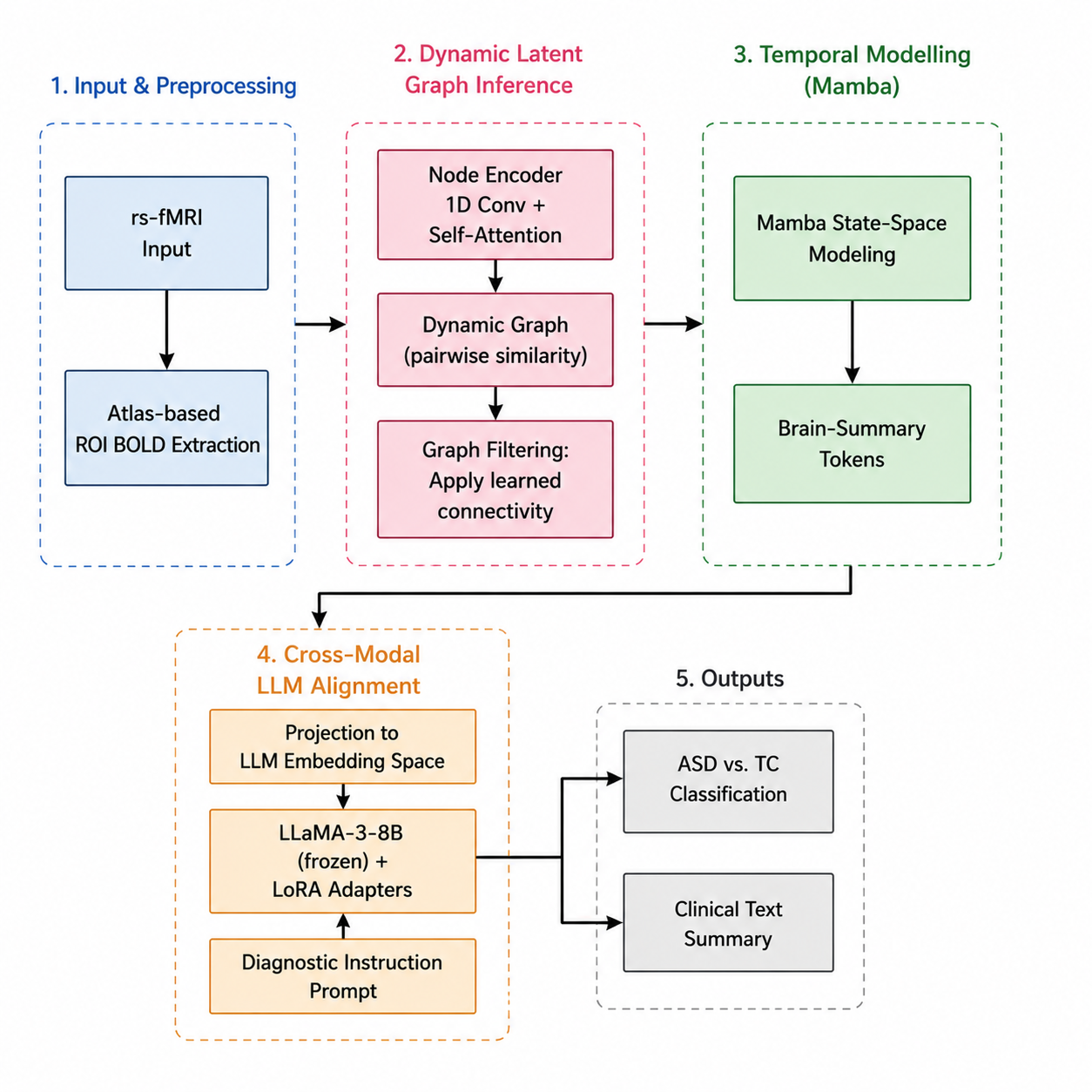}
    \caption{Overview of the NeuroMambaLLM pipeline.}
    \label{fig:pipeline}
\end{figure}

To bridge the gap between dynamic neurobiological processes and computational feasibility, we propose NeuroMambaLLM, a framework that shifts the paradigm from Static Graph-to-Text toward Dynamic Latent Signal-to-Text modelling. Our approach integrates adaptive latent graph inference, efficient state-space sequence modelling, and large language model reasoning within a unified architecture. Along with graph neural networks, diffusion-based generative models have similarly been applied for latent representation in high-dimensional systems\cite{Torabi2026PIGNN, torabi2025diffusion_microring}. First, rather than relying exclusively on precomputed correlation matrices or fixed structural priors as in approaches such as Graph Diffusion Autoregression (GDAR) \cite{gdar2025}, we employ Adaptive Latent Graph Inference (LGI), a differentiable module that learns time-resolved functional connectivity directly from raw Blood-Oxygen-Level-Dependent (BOLD) signals. By projecting BOLD activity into a latent interaction space, LGI captures non-linear dependencies and transient functional reorganization that are poorly represented by static correlation metrics \cite{Lu2023LatentGI}. Second, we adopt the Mamba selective state-space architecture to model long-range temporal dependencies while avoiding the high computational cost associated with Transformers \cite{gu2023mamba}. Mamba is a state-space sequence modelling architecture that processes signals in linear time and adaptively selects relevant information at each time step. It is particularly well-suited for long and noisy fMRI time series where both efficiency and robustness are essential. Unlike recurrent models that suffer from forgetting or attention-based models that incur quadratic memory growth, Mamba enables linear-time sequence modelling through a hardware-aware selective scan mechanism. While recent studies have begun applying Mamba to physiological signals, these approaches typically employ it as a standalone feature extractor rather than as part of an end-to-end reasoning framework. Finally, we align the resulting dynamic graph representations with the embedding space of a large language model (LLaMA-3-8B), enabling high-level semantic reasoning over temporally specific neural dynamics. In clinical settings, a black-box prediction of ASD is insufficient; interpretability and contextual reasoning are essential. By embedding time-resolved brain graph representations into the latent space of the language model, NeuroMambaLLM enables diagnostic classification while grounding its outputs in causally relevant temporal patterns (Fig.~\ref{fig:pipeline}). 

Our contributions are summarized as follows:
\begin{enumerate}
\item \textbf{Dynamic Functional Graph Learning:} We introduce an end-to-end dynamic graph encoder that learns time-resolved functional connectivity directly from raw BOLD signals using one-dimensional convolution and self-attention, capturing transient neural interactions missed by static analyses.
\item \textbf{Efficient Temporal Modelling with Mamba:} We employ the Mamba selective state-space architecture to efficiently model long-range temporal dependencies in fMRI data while adaptively suppressing motion and physiological noise without the quadratic cost of Transformers.
\item \textbf{Large Language Model Reasoning:} We propose a framework that projects the brain graphs into the embedding space of a frozen LLaMA-3-8B using lightweight LoRA modules. This enables joint diagnostic classification and language-based reasoning, producing interpretable and clinically meaningful textual reports.
\end{enumerate}

\section{Related Work}

\subsection{Large Language Models for Brain Connectivity}
The integration of LLMs with neuroimaging is a new and rapidly expanding field. Early efforts focused on translating text reports to images, but recent works like FCN-LLM \cite{fcnllm2025} and BrainLLM \cite{zhou2024brainllm} have attempted to invert this process, using LLMs to interpret functional connectivity networks (FCNs). These models typically employ a projection layer (e.g., MLP or Q-Former) to map flattened upper-triangular correlation matrices into the LLM's token space. While innovative, these methods invariably rely on static input matrices. By averaging out temporal dynamics, they discard the richness of time, which limits the LLM's reasoning capabilities to average biomarkers rather than specific temporal events.

\subsection{Dynamic Functional Connectivity (dFC) Learning}
To overcome the limitations of static analysis, deep learning approaches have increasingly targeted Dynamic Functional Connectivity (dFC). Spatiotemporal Graph Convolutional Networks (ST-GCNs) \cite{gadgil2020spatio, kim2021learning} and Transformer-based models like the Brain Network State Transformer \cite{bnst2025} have shown promise in capturing time-varying dependencies. However, these methods often rely on sliding-window techniques, which introduce hyperparameter sensitivity (i.e., window size selection). Furthermore, methods such as Graph Diffusion Autoregression (GDAR) \cite{gdar2025} rely on fixed structural priors, potentially missing functional reorganization that deviates from anatomical connections. These dFC models operate as black boxes, which offer high classification accuracy but they lack the linguistic interpretability required for clinical trust.

\subsection{State Space Models (Mamba) in Medical Imaging}
The recently introduced Mamba architecture \cite{gu2023mamba}, based on Selective State Space Models (SSMs), has revolutionized long-sequence modelling by achieving linear computational complexity. This has led to its rapid adoption in medical imaging, including segmentation (SegMamba \cite{xing2024segmamba}, U-Mamba \cite{ma2024umamba}) and classification. In the domain of biosignals, and Mamba-CAM-Sleep \cite{mambasleep2025} utilize Mamba to process EEG and fMRI time series, respectively. However, these works primarily use Mamba as a standalone backbone for feature extraction. Our work is the first to integrate Latent Graph Learning with Mamba-LLM alignment, utilizing the state-space model not just for classification, but to create a compressed, time-resolved context for high-level diagnostic reasoning.
\section{Methodology}
\label{sec:methodology}

Our framework models brain activity as a sequence of dynamically evolving functional graphs. The architecture consists of three coupled stages: (1) Dynamic Latent Graph Inference, (2) Selective State-Space Temporal Modelling, and (3) Cross-Modal Alignment with a Large Language Model.

\subsection{Theoretical Formulation}

\subsubsection{Dynamic Latent Graph Inference} 

Let the input fMRI batch be denoted by $\mathbf{X} \in \mathbb{R}^{B \times T \times N}$, where $B$ represents the batch size, $T$ is the number of time points in the scan, and $N$ is the number of brain Regions of Interest (ROIs). At each time step $t \in [1, T]$, the instantaneous brain activity is represented by the vector $\mathbf{x}_t \in \mathbb{R}^{N}$, where the scalar $x_{t,i}$ corresponds to the Blood-Oxygen-Level-Dependent (BOLD) signal of the $i$-th ROI. We model whole-brain activity at each time point as a dynamic functional graph
$\mathcal{G}_t = (\mathcal{V}, \mathcal{E}_t)$, where $\mathcal{V}$ is the set of
$N$ Regions of Interest (ROIs) and $\mathcal{E}_t$ denotes time-varying functional
connections between them. Unlike static functional connectivity, the graph
structure is inferred dynamically from the raw fMRI signal at each time step. Each scalar BOLD signal is
mapped into a latent feature space using a shared learnable node encoder
$\phi(\cdot)$, implemented using one-dimensional convolution followed by
self-attention to capture local temporal patterns and inter-regional dependencies.
The latent embedding of node $i$ at time $t$ is given by:
\begin{equation}
    \mathbf{h}_{t,i} = \phi(x_{t,i}) \in \mathbb{R}^{d_{\text{lat}}},
\end{equation}
where $d_{\text{lat}}$ denotes the latent embedding dimension. Stacking all node
embeddings yield the latent node matrix
$\mathbf{H}_t = [\mathbf{h}_{t,1}, \dots, \mathbf{h}_{t,N}]^\top \in \mathbb{R}^{N \times d_{\text{lat}}}$. Dynamic functional connectivity is inferred directly from these latent
representations by computing pairwise similarity between node embeddings. The
time-varying adjacency matrix $\mathbf{G}_t \in \mathbb{R}^{N \times N}$ is defined
as:
\begin{equation}
    G_{t,ij} = \frac{\mathbf{h}_{t,i}^\top \mathbf{h}_{t,j}}{\sqrt{d_{\text{lat}}}},
\end{equation}
which captures instantaneous functional coupling between ROIs. This graph is a latent, task-driven interaction model inferred from fMRI time series \cite{Razmara2025TaskfMRI}, rather than explicit neurobiological functional connectivity. This formulation avoids heuristic sliding windows and allows connectivity to evolve continuously over time. The inferred connectivity is then used to spatially contextualize the brain
activity:
\begin{equation}
    \tilde{\mathbf{x}}_t = \mathbf{G}_t \mathbf{x}_t,
\end{equation}
where $\tilde{\mathbf{x}}_t \in \mathbb{R}^{N}$ represents the dynamically filtered
brain activity. Let $\tilde{\mathbf{X}} = [\tilde{\mathbf{x}}_1, \dots, \tilde{\mathbf{x}}_T]$ denote the
sequence of graph-filtered brain states.

\subsubsection{Selective State-Space Temporal Modelling}

Temporal dependencies are modelled using a Selective State-Space Model
(SSM) based on the Mamba architecture. The model maintains a latent temporal state
$\mathbf{s}_t \in \mathbb{R}^{d_h}$, initialized as $\mathbf{s}_0 = \mathbf{0}$, and
updated recursively according to:
\begin{equation}
    \mathbf{s}_t = \mathbf{A}_t \mathbf{s}_{t-1} + \mathbf{B}_t \tilde{\mathbf{x}}_t,
\end{equation}
where $\mathbf{A}_t \in \mathbb{R}^{d_h \times d_h}$ is the state transition matrix
and $\mathbf{B}_t \in \mathbb{R}^{d_h \times N}$ is the input projection matrix at time $t$. Here, $d_h$ denotes the dimensionality of the latent temporal state maintained by the
state-space model. In Mamba, the parameters $\mathbf{A}_t$ and $\mathbf{B}_t$ are not fixed but are
generated as deterministic functions of the current input
$\tilde{\mathbf{x}}_t$, enabling input-dependent state evolution. This selective
parameterization allows the model to attenuate noise-dominated time points while
preserving temporally localized neural patterns. The formulation supports linear-time
sequence modelling with respect to $T$, making it well-suited for long fMRI sequences.

\subsubsection{Cross-Modal LLM Alignment}

Let $\mathbf{S} = [\mathbf{s}_1, \dots, \mathbf{s}_T] \in \mathbb{R}^{T \times d_h}$
denote the sequence of latent temporal states. To interface with a Large Language
Model (LLM), this variable-length sequence is compressed into a fixed set of
brain-summary tokens:
\begin{equation}
    \mathbf{Z}_{\text{brain}} \in \mathbb{R}^{K \times d_k},
\end{equation}

\subsection{Data Extraction and Functional Connectivity}
The fMRI preprocessing requires accurate atlas registration and BOLD signal extraction. We convert raw volumetric fMRI scans into graph-structured representations using a standardized anatomical atlas. We align all scans to MNI space to ensure spatial consistency and extract the mean BOLD signal from each Region of Interest (ROI), producing $N$ representative time-series nodes. Traditional functional connectivity methods, such as Pearson Correlation Coefficient (PCC), capture static linear relationships between ROI pairs \cite{vidaurre2021brain}, while extensions like Tangent Pearson Embedding (TPE) model non-linear structure \cite{kim2021learning}. However, these approaches average out temporal dynamics. In contrast, we infer time-resolved latent connectivity directly from ROI-level BOLD signals to explicitly model dynamic brain interactions.

\subsection{Mamba Selective State-Space Modelling}
We model long-range temporal dependencies in fMRI sequences using the Mamba architecture based on Selective State Space Models (SSMs) \cite{gu2023mamba}. In our framework, individual Mamba blocks process each ROI's time-series to extract informative temporal features, emphasizing task-relevant neural patterns while suppressing noise and motion-related artifacts. The resulting features are then aggregated into a representation of the brain’s dynamic state, which we subsequently align with the language model.

\subsection{LLM Reasoning}

The final stage of NeuroMambaLLM aligns compressed dynamic brain representations with the semantic space of the LLM. A projection layer maps the Mamba-encoded temporal embeddings into the LLM token space. The LLM performs diagnostic inference and generates clinically meaningful textual summaries. Its input consists of a task-specific textual input concatenated with projected brain-state tokens. The base LLM is kept fixed, while small LoRA modules are trained together with the projection and brain encoders. This lightweight training allows the model to adapt to neuroimaging data without retraining the full LLM, producing both ASD classification and clear text explanations grounded in brain dynamics \cite{fcnllm2025}.

\section{Experiments}
\subsection{Datasets}

We conduct experiments using resting-state functional MRI (rs-fMRI) data from the Autism Brain Imaging Data Exchange (ABIDE I) \cite{diMartino2014ABIDE}, a large multi-site initiative designed to advance research on the neural mechanisms of autism. ABIDE aggregates data collected across multiple international sites, resulting in a heterogeneous dataset that reflects real-world variability in scanner hardware, acquisition protocols, and scan durations. The dataset comprises individuals diagnosed with Autism Spectrum Disorder (ASD) and Typically Developing Controls (TC). Subjects are randomly split into 80\% for training and 20\% for testing at the subject level. We do not use phenotypic or demographic information. Instead, the proposed model relies exclusively on fMRI time-series data. Data are preprocessed to reduce noise and ensure spatial consistency, including correction for head motion, slice timing differences, intensity variations, and alignment to the standard space. Brain activity is represented at the regional level using an atlas-based definition of $N$ regions of interest ($N$= 39 ROIs). For each ROI, we extract the mean BOLD signal over time and normalize each time series to zero mean and unit variance. Each fMRI scan is truncated to a fixed length of $T$ = 100 time points. These ROI-level BOLD time series serve as input to the model.
\subsection{Experimental Setup}

All experiments were implemented in PyTorch \footnote{ \url{https://github.com/hamedajorlou/NeuroMambaLLM}}. The model takes raw ROI-level resting-state fMRI BOLD time series as input and performs end-to-end learning without constructing handcrafted functional connectivity matrices. The dynamic graph encoder applies grouped one-dimensional temporal convolutions with kernel size 3 and ReLU activation to extract local temporal features independently for each ROI, followed by a self-attention mechanism that infers time-resolved functional connectivity in a latent embedding space of dimension 128. The resulting graph-aware representations are processed by a temporal encoder with two layers and four attention heads to capture long-range dependencies across the fMRI sequence. For cross-modal reasoning, we employ the Large Language Model Meta AI (LLaMA-3-8B) as the language backbone. The LLM is loaded using 4-bit quantization to reduce memory usage, and all base model parameters remain frozen during training. Parameter-efficient fine-tuning is achieved by inserting Low-Rank Adaptation (LoRA) modules with rank $r=16$, scaling factor $\alpha=32$, and dropout 0.1 into the attention layers. A fixed set of learnable brain-summary tokens compresses the temporal representations and projects them into the LLM embedding space, where they are concatenated with a diagnostic instruction prompt. Training is performed using mixed-precision arithmetic and gradient accumulation on a single A100 GPU, optimized with the Adam optimizer at a learning rate of $1\times10^{-4}$ for up to 10 epochs.

\subsection{Evaluation and Results}

The model achieves an overall classification accuracy of $0.7212 \pm 0.0098$. The proposed method reaches a precision of $0.8022 \pm 0.0401$, indicating reliable identification of autism cases among predicted positives, while the recall reaches $0.6102 \pm 0.0602$, reflecting balanced sensitivity under dataset heterogeneity and limited sample size. The resulting F1-score of $0.6931 \pm 0.0560$ demonstrates a favourable trade-off between precision and recall.  

\begin{table*}[]
\centering
\caption{Comparison of different methods on the ABIDE dataset.}
\label{tab:comparison}
\begin{tabular}{l l c c c c}
\toprule
\textbf{Method} & \textbf{Description} & \textbf{Accuracy} & \textbf{Precision} & \textbf{Recall} & \textbf{F1-score} \\
\midrule

BoLT \cite{bolt} 
& Brain-oriented long-range transformer for temporal fMRI modelling 
& \underline{\underline{0.7128}} & 0.7132 & \textbf{0.6485} & \underline{\underline{0.6799}} \\

SwinT \cite{swint} 
& Hierarchical spatiotemporal transformer for fMRI 
& 0.6859 & 0.6827 & 0.6075 & 0.6427 \\

CNN--LSTM \cite{cnn_lstm} 
& CNN-based spatial encoder with long short-term memory 
& 0.6549 & 0.6479 & 0.5731 & 0.6081 \\

BrainGNN \cite{braingnn} 
& Graph neural network on static functional connectivity 
& \underline{0.6931} & -- & -- & 0.6389 \\

BrainNPT \cite{brainnpt} 
& Neural process transformer for brain representation learning 
& 0.6810 & -- & -- & 0.6453 \\

PTGB \cite{ptgb} 
& Prompt-tuned graph-based brain foundation model 
& 0.6709 & -- & -- & 0.6520 \\

CINP \cite{cinp} 
& Contrastive instance-wise neural projection model 
& 0.6760 & -- & -- & 0.6474 \\

BrainMass \cite{brainmass} 
& Masked modelling of multivariate brain signals 
& 0.6926 & -- & -- & 0.6398 \\

\midrule
\textbf{NeuroMambaLLM (Proposed)} 
& \textbf{Dynamic latent brain graph with LLM-guided reasoning} 
& \textbf{0.7212} & \textbf{0.8022} & \underline{0.6102} & \textbf{0.6931} \\

Static MambaLLM
& LLM-enhanced Mamba state-space model without graph dynamics 
& 0.6650 & \underline{0.7410} & 0.5980 & 0.6620 \\

Frozen MambaLLM 
& Frozen LLM backbone without LoRA adaptation 
& 0.6820 & \underline{\underline{0.7630}} & \underline{\underline{0.6120}} & \underline{0.6790} \\

\bottomrule
\addlinespace
\multicolumn{6}{l}{\footnotesize\textit{Note: \textbf{Bold} values indicate the best result in each column; \underline{\underline{double}} and \underline{single} underlines denote the second and third best results, respectively.}}
\end{tabular}
\end{table*}

Table~\ref{tab:comparison} compares the proposed framework with recent methods on the ABIDE dataset.
Among different approaches, NeuroMambaLLM achieves higher accuracy and F1-score. The last three rows of Table~\ref{tab:comparison} analyze the impact of dynamic graph modelling and parameter-efficient adaptation. Replacing the dynamic graph with a static formulation (Static MambaLLM) degrades performance, highlighting the importance of adaptive graph inference. In addition, freezing the LLM parameters (Frozen MambaLLM) reduces accuracy and F1-score compared to LoRA adaptation, confirming the value of lightweight fine-tuning.

\subsection{Ablation Study}

\begin{figure}[!ht]
    \centering
    \includegraphics[width=0.5\textwidth]{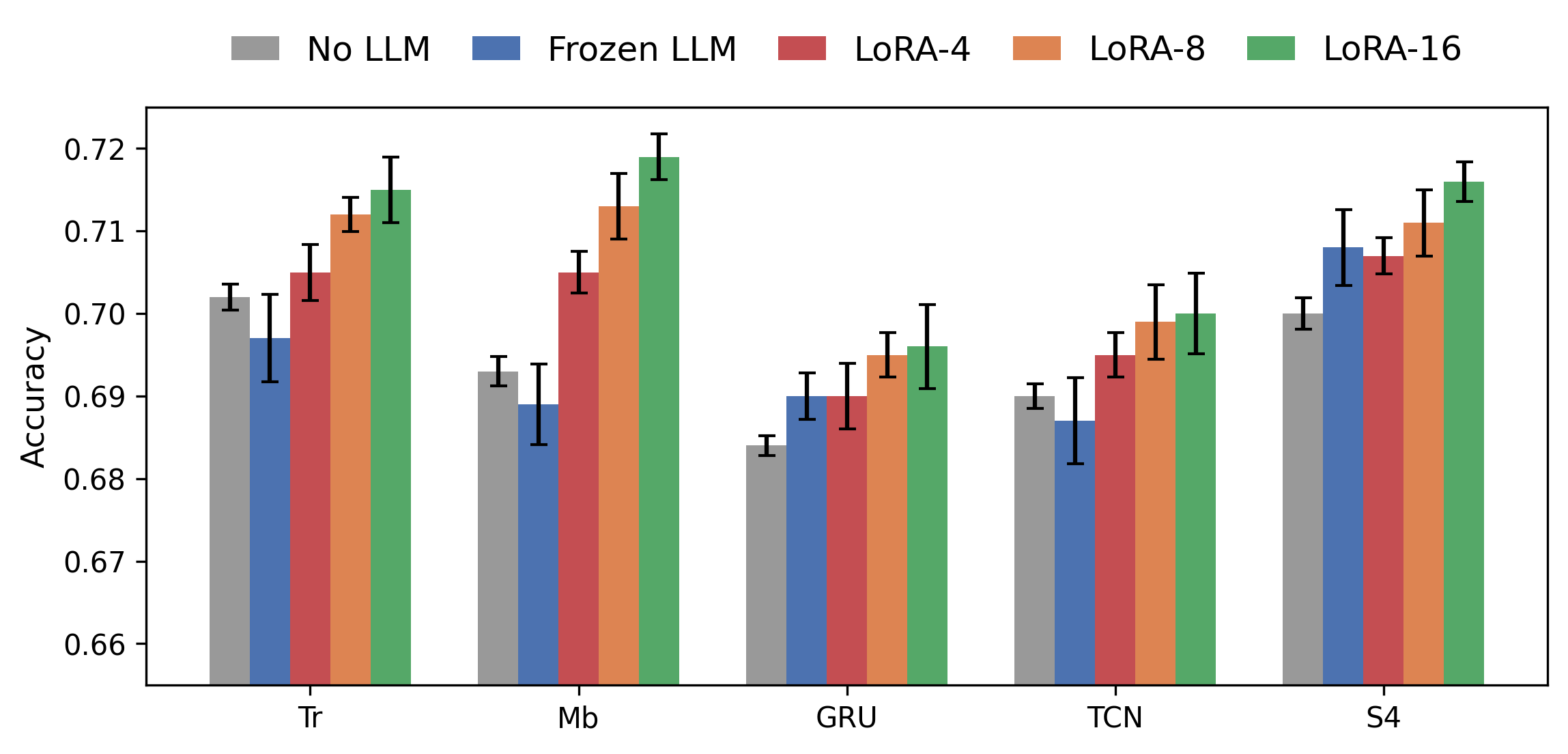}
    \caption{
Ablation results. Comparison of temporal modelling backbones and the effect of integrating an LLM with frozen and LoRA-based adaptation.
    }
    \label{fig:mamba4}
\end{figure}

We conduct an ablation study to evaluate key components of the proposed framework.
All models share the same input features, graph construction, and training protocol, and differ only in the temporal modelling backbone or LLM adaptation strategy. We replace the Mamba-based temporal module with gated recurrent units (GRU), temporal convolutional networks (TCN), Transformers (Tr), and structured state-space models (S4) to compare different approaches for modelling fMRI time series over time, and observe that Mamba (Mb) provides strong performance (Fig.~\ref{fig:mamba4}). We further evaluate the impact of integrating a large language model (LLM) and frozen versus LoRA-based adaptation with different ranks ($r=4,8,16$), showing that lightweight fine-tuning improves performance. The reported performance improvements are consistent across multiple random train–test splits, with low variance as indicated by the error bars.

\subsection{LLM Reasoning}

\begin{figure}[!ht]
    \centering
    \includegraphics[width=0.35\textwidth]{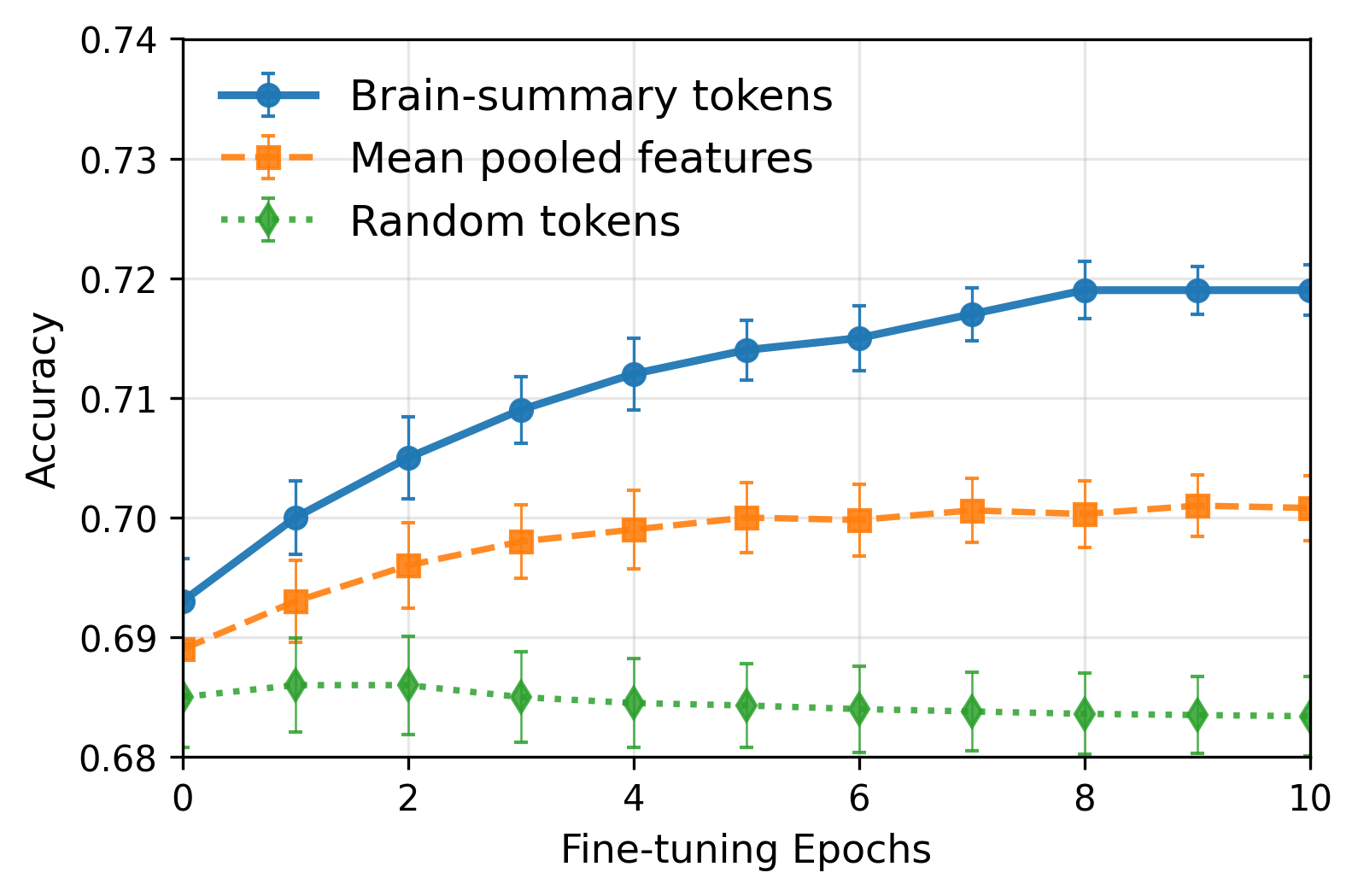}
    \caption{
Effect of brain–LLM alignment during fine-tuning. Structured brain-summary tokens improve performance, while mean pooling offers limited gains and random tokens provide no benefit.
    }
    \label{fig:llm_alignment}
\end{figure}

In addition to classification, the LLM is used to generate short, clinically interpretable summaries from learned dynamic brain representations. The projected brain tokens are concatenated with a simple diagnostic prompt, allowing the LLM to reason over regional activations and connectivity patterns and express them in natural language. To reduce hallucinations, we provide the LLM with a predefined description schema that constrains the generated text to connectivity statistics. This enables the model to explain why a subject is classified as ASD or control, rather than only providing a label and an unrelated text. An illustrative example of the prompt and generated output is shown in Fig.~\ref{fig:llm_prompt}. Fig.~\ref{fig:llm_alignment} shows that structured brain–LLM alignment using brain-summary tokens leads to steady accuracy improvements, whereas mean pooling yields limited gains and random token inputs provide no benefit. This indicates that LLM performance depends on meaningful alignment rather than token injection alone.

\begin{figure}[!ht]
\centering

\begin{tcolorbox}[promptbox, title={LLM Prompt}, width=\linewidth]
\footnotesize
Generate a descriptive summary of the observed fMRI patterns based on the Mamba analysis. The summary should support exploratory analysis and expert interpretation and should not be considered a clinical diagnosis.
\end{tcolorbox}

\begin{tcolorbox}[promptbox, title={LLM Output}, width=\linewidth]
\footnotesize
The model identifies atypical temporal interactions involving frontal and temporal regions, with reduced long-range integration and increased local connectivity. Such patterns have been reported in autism spectrum disorder, but this observation is intended for exploratory analysis only. Classification leaning toward ASD (58.0\% confidence).
\end{tcolorbox}

\caption{Example LLM prompt and generated clinical response.}
\label{fig:llm_prompt}
\end{figure}

\subsection{Neurobiological Interpretation}

\begin{figure}[H]
    \centering
    \includegraphics[width=0.5\textwidth]{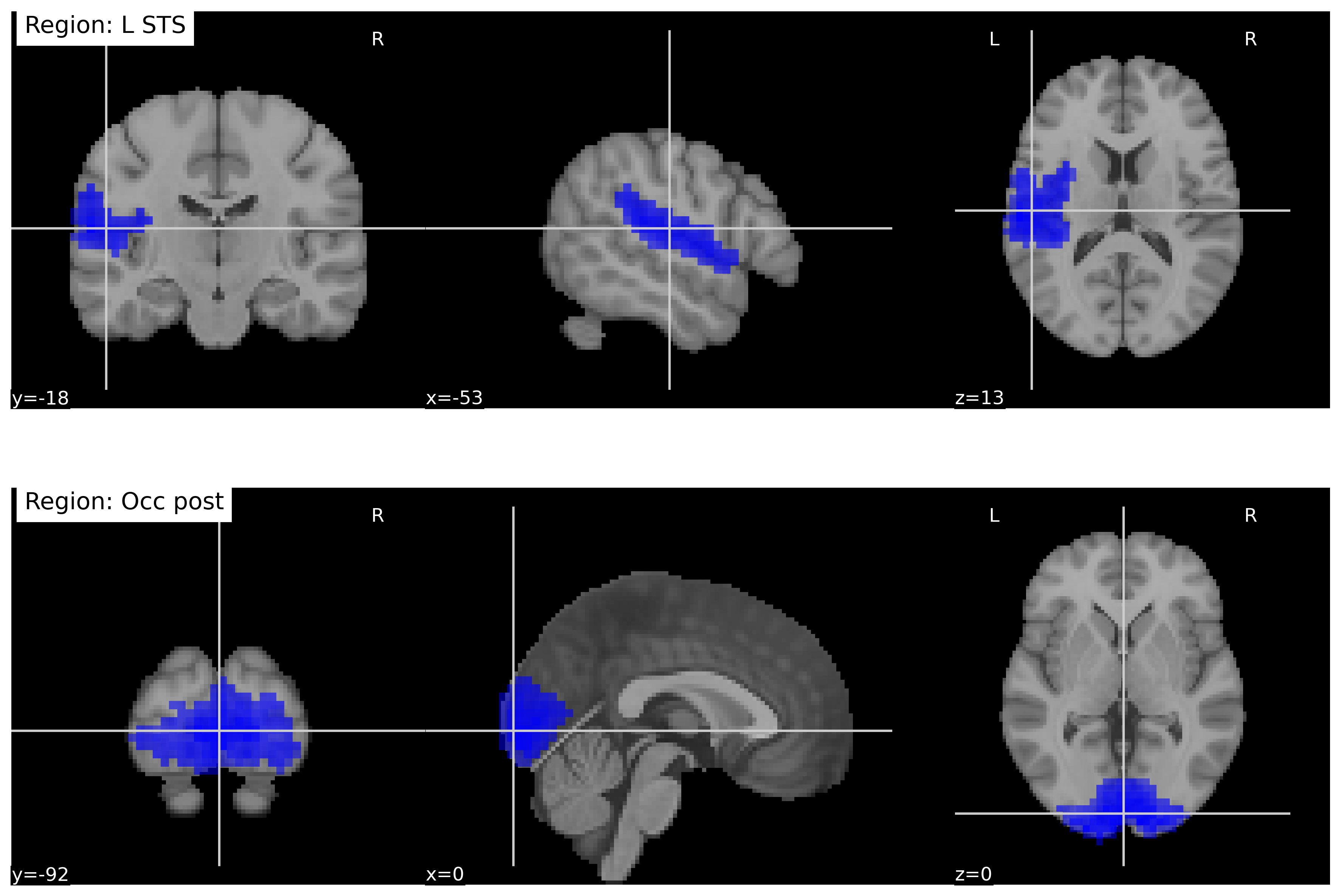}
    \caption{Regional brain activations based on NeuroMambaLLM Results. Prominent involvement of the left superior temporal sulcus (STS) and posterior occipital regions highlights circuits related to social perception/audiovisual integration and early visual-sensory processing.}
    \label{fig:activation}
\end{figure}

The learned regional saliency and connectivity patterns provide insight into the neurobiological mechanisms captured by the proposed model. Overall, the extracted features are consistent with known alterations in sensory, temporal, and associative brain systems in ASD \cite{belmonte2004autism}. Fig.~\ref{fig:activation} shows that the model assigns high importance to the left superior temporal sulcus (STS) and posterior occipital cortex, regions involved in social perception, language, and audiovisual integration. Atypical structure and function of the STS and posterior temporal--occipital regions have been widely reported in ASD and are closely linked to deficits in social communication and sensory processing \cite{pelphrey2004perception}. 

\begin{figure}[H]
\vspace{-4mm}
    \centering
    \includegraphics[
        width=0.5\textwidth,
        trim=10 150 10 100,
        clip
    ]{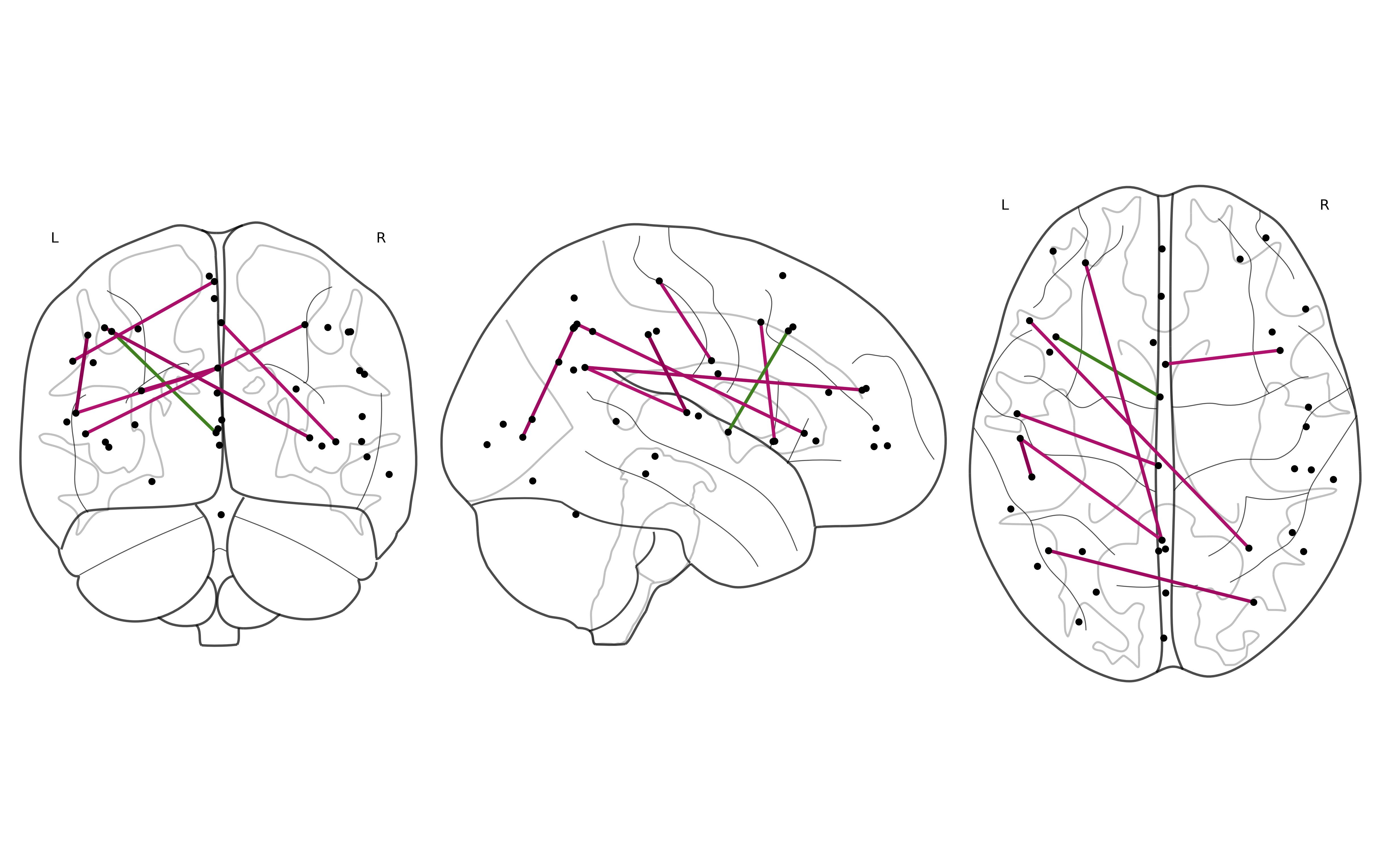}

    \caption{Most informative functional connectivity patterns projected onto the cortical surface. From left to right: left lateral, medial, and right lateral views. The model highlights predominantly intra-hemispheric and long-range fronto-temporal connections, with left-hemisphere dominance.}
    \label{fig:surface_connectivity}
\end{figure}

At the network level, Fig.~\ref{fig:surface_connectivity} reveals dominant intra-hemispheric and long-range frontotemporal interactions, with a pronounced left-lateralized pattern. This finding is consistent with evidence of altered hemispheric specialization and disrupted functional integration between frontal and temporal regions in ASD \cite{just2007functional}. 

\begin{figure}[h]
    \centering
    \includegraphics[
        width=0.4\textwidth,
        trim=90 30 90 10,
        clip
    ]{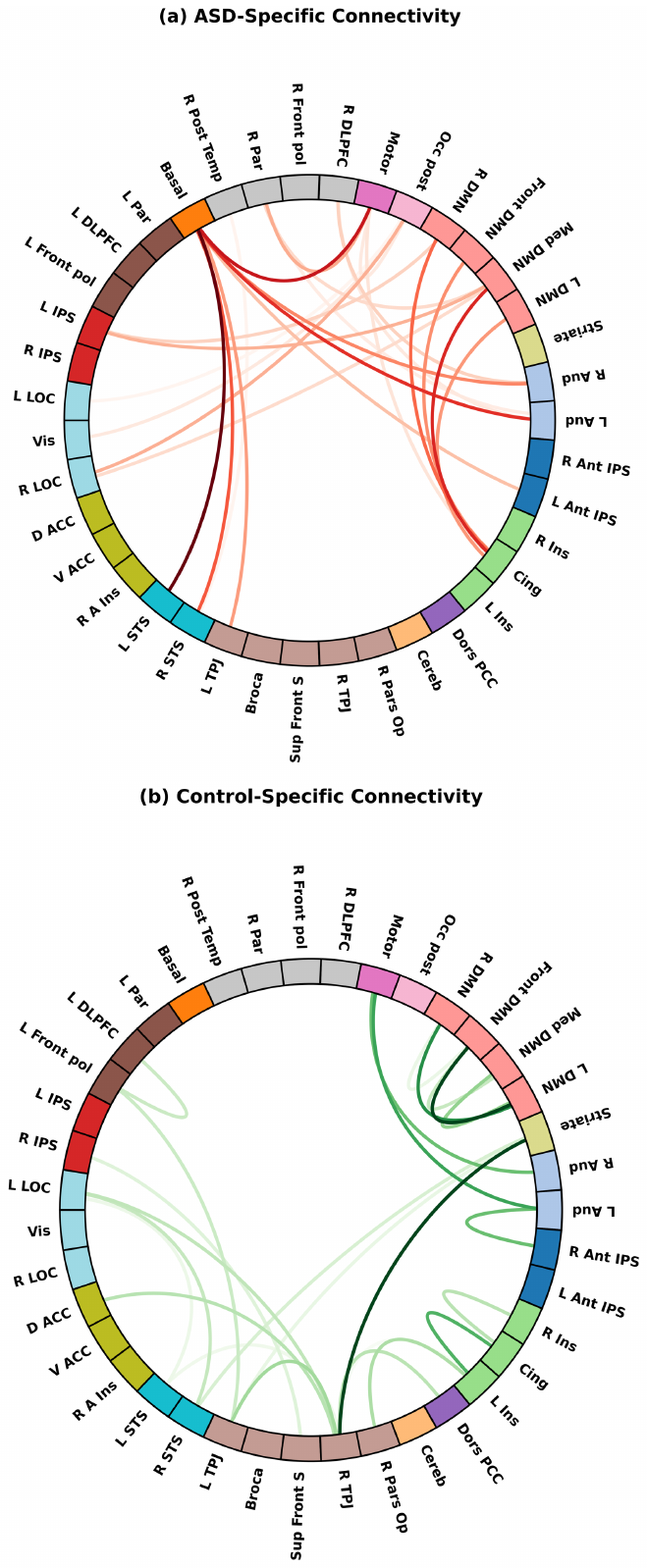}
\caption{Group-specific functional connectivity patterns. (a) Autism Spectrum Disorder (ASD) connectivity (red edges). (b) Typically Developing Controls (TC) connectivity (green edges). Nodes correspond to anatomically defined Regions of Interest (ROIs), labelled by standard abbreviations and hemisphere, and organized according to functional systems: visual (VIS), somatomotor (SM), dorsal attention (DAN), ventral attention/salience (VAN), limbic (LIM), frontoparietal control (FPN), and default mode network (DMN). Edge thickness indicates relative connection importance.}

    \label{fig:connectome}
\end{figure}

The circular connectome in Fig.~\ref{fig:connectome} illustrates the connectivity patterns, where ASD-specific connections appear more spatially concentrated and selective, particularly within temporal and associative regions, whereas control-specific connectivity shows broader and more distributed inter-regional integration. The ASD edges appear stronger and more region-focused, while control networks retain wider cross-network coupling. This supports the view that ASD reflects an imbalance between local specialization and global functional integration \cite{courchesne2005local}. 

\section{Conclusion}
We introduced NeuroMambaLLM, a framework that learns dynamic functional connectivity from fMRI time series and combines temporal modelling with language model reasoning. By avoiding fixed connectivity assumptions, the method captures time-varying brain interactions relevant for autism classification. Experiments on the ABIDE dataset demonstrate competitive performance and connectivity patterns consistent with known neurobiological findings. Future work will assess generalization across additional datasets and imaging sites. LLM-generated explanations may be affected by limitations such as hallucination or overgeneralization; future work will explore constrained prompting and confidence-aware decoding. Importantly, the language model is not introduced to maximize classification accuracy, but to support structured interpretation of dynamic brain patterns in a form that is accessible to domain experts. Compared to classical models that provide only labels or scores, this approach offers a more complete analysis of temporal and connectivity information, supporting expert reasoning without replacing clinical judgment. Overall, linking dynamic brain representations with language-based reasoning enhances interpretability beyond standard black-box models and opens directions for richer clinical reasoning.

\bibliographystyle{IEEEtran}
\bibliography{ref}

\end{document}